\documentclass[12pt,a4paper]{article}
\usepackage{epsfig}
\begin{document}
\title{The new charged gauge boson $W'$ and the subprocess $eq\rightarrow\nu q'$ at $e^{+}e^{-}$ and $ep$ colliders}
\author{Chong-Xing Yue, Li Ding, Wei Ma\\
{\small Department of Physics, Liaoning  Normal University, Dalian
116029, P. R. China}
\thanks{E-mail:cxyue@lnnu.edu.cn}}
\date{\today}

\maketitle
\begin{abstract}

\vspace{1cm}

In the framework of the little Higgs models and the three-site
Higgsless model, we discuss the contributions of the new charged
gauge boson $W'$ to the process $eq\rightarrow\nu q'$  and the
possibility of detecting $W'$ via this process in future high energy
linear $e^{+}e^{-}$ collider $(ILC)$ and $ep$ collider $(THERA)$
experiments. Our numerical results show that the process
$eq\rightarrow\nu q'$ is rather sensitive to the coupling $W'ff'$
and one can use this process to distinguish different new physics
models in future $ILC$ and $THERA$ experiments.
\end{abstract}

\hspace{0.4cm}$PACS$ $number$:12. 60{\it Cn}, 14. 70{\it Pw}, 14.
65.{\it Bt}
\newpage
\noindent{\bf 1. Introduction}

Although any new charged gauge boson, generally called $W'$, is not
found yet experimentally, its existence is now a relatively common
prediction which results from many new physics scenarios. For
example, little Higgs models [1], Higgsless models [2],
non-commuting extended technicolor [3], and Randall-Sundrum model
with bulk gauge fields [4] give examples where extension of gauge
group lead to appearing of $W'$. If one of these new particles is
discovered, it would represent irrefutable proof of new physics,
most likely that the gauge group of the standard model $(SM)$ must
be extended. Thus, search for extra gauge boson $W'$ provides a
common tool in quest for new physics at the next generation collider
experiments [5].

Although the extra gauge boson $W'$ is not discovered yet there are
experimental limits on its mass. The indirect limits can be placed
on the existence of $W'$ through indirect searches based on the
deviations from the $SM$, which can be obtained in precision
electroweak measurements [6, 7]. Indirect searches for $W'$ being
extracted from leptonic and semileptonic decays and also from
cosmological and astrophysical data give very wide range for upper
limits on $W'$ mass varying from 549$GeV$ up to 23$TeV$ [8]. The
direct limits on $W'$ mass are based on hypothesis of purely right
or left-handed interacting $W'$ with $SM$-like coupling constants
[9]. At hadron colliders, the limits can be obtained by considering
its direct production via the $Drell-Yan$ process and its subsequent
decay to lepton pairs or hadronic jets. Present bounds from
measurements at the Tevatron collider exclude low $ W'$ mass,
$M_{W'}>720GeV$ [10]. The $CERN$ Large Hadron Collider $(LHC)$ is
expected to be able to discover $W'$ up to mass of $\approx5.9TeV$
[11].

So far, there are some studies of indirect searches for $W'$ boson
at high energy colliders. For example, Ref.[6] has examined the
sensitivity of the process $e^{+}e^{-}\rightarrow
\nu\bar{\nu}\gamma$ to the mass of $W'$ boson and found that this
process is sensitive to $W'$'s mass up to several $TeV$. Ref.[7]
further studied the sensitivity of the process $e\gamma\rightarrow
\nu q +X$ to $W'$ boson and compared with the process
$e^{+}e^{-}\rightarrow \nu\bar{\nu}\gamma$, which find that, in many
cases, this process is more sensitive to $W'$ boson than that of the
process $e^{+}e^{-}\rightarrow \nu\bar{\nu}\gamma$. Recently,
Ref.[12] has explored the capability of the $LHC$ to determine the
$W'$ coupling helicity at low integrated luminosities in the
$l+E_{T}^{miss}$ discovery channel and Ref.[13] has further studied
the process $e^{+}e^{-}\rightarrow \nu\bar{\nu}\gamma$ in the
context of the little Higgs model. Ref.[14] has studied the
possibility of detecting the  $W'$ boson predicted by the three-site
Higgsless model via the processes $pp\rightarrow W'\rightarrow WZ$
and $pp\rightarrow W'jj\rightarrow WZjj$ at the upcoming $LHC$. In
this paper, we will calculate the corrections of the gauge boson
$W'$ to the process $eq\rightarrow \nu q'$ in different extensions
of the $SM$ and see whether this process can be used to distinguish
different new physics models in future high energy collider
experiments.

Little Higgs theory is proposed as an alternative solution to the
hierarchy problem of the $SM$, which provides a possible kind of
electroweak symmetry breaking $(EWSB)$ mechanism accomplished by a
naturally light Higgs boson [1]. In general, this kind of models
predict the existence of the pure left-handed charged gauge boson
$W'$, which has the $SM$-like couplings to ordinary particles. In
this paper, we will first consider the process $eq\rightarrow \nu
q'$ in this kind of models. The second kind of models are the
Higgsless models, which have been proven to be viable alternative to
the $SM$ and supersymmetric models in describing the breaking of the
electroweak symmetry [15]. The three-site Higgsless model [16] is
one of the simplest and phenomenologically viable models and has all
essential features of the Higgsless models. Thus we will consider
the contributions of the charged $KK$ gauge boson $W'$ predicted by
the three-site Higgsless model to the process $eq\rightarrow \nu
q'$.

Section 2 of this paper contains the elemental formula, which are
related to our calculation. Based on the structure of the extended
electroweak gauge group, the little Higgs models can be divided into
two classes [17, 18]: the product group models and the simple group
models. The littlest Higgs model $(LH)$ [19] and the $SU(3)$ simple
group model [18, 20] are the simple examples of these two kinds of
little Higgs models, respectively. The contributions of these two
models to the process $eq\rightarrow \nu q'$ are considered and the
relevant phenomenology analysis in future high energy linear
$e^{+}e^{-}$ collider $(ILC)$ [21] and $ep$ collider $(THERA)$ [22]
are given in section 3. Section 4 gives our numerical results
obtained in the framework of three-site Higgsless model. In the last
section the summary and discussion are given.

\newpage

\noindent{\bf 2. The relevant formula about our calculation}

To consider the $W'$ contributions to the process $eq\rightarrow \nu
q'$ in different new physics scenarios, we write down the lowest
dimension effective Lagrangian of $W'$ interactions to ordinary
fermions in most general form (possible higher dimension effective
operators are not taken into account in our numerical calculation):
 \begin{equation}
\pounds=\frac{e}{\sqrt{2}S_{W}}V_{ij}\bar{f_{i}}\gamma^{\mu}(g_{L}P_{L}+g_{R}P_{R})f_{j}W'_{\mu}+h.c.
,
\end{equation}
where $S_{W}=\sin\theta_{W}$ ($\theta_{W}$ is the Weinberg angle),
$V_{ij}$ is the $CKM$ matrix element, and
$P_{L(R)}=(1\mp\gamma_{5})/2$ is the left-(right-) handed projection
operator. In the $SM$ case, the coupling constant $g_{L}$ is equal
to one and $g_{R}$ is equal to zero.

The production cross section $\hat{\sigma}(\hat{s})$ of the process
$e(P_{1})+q(P_{2})\rightarrow \nu(P_{3})+q'(P_{4})$ contributed by
the $SM$ gauge boson $W$ and the new charged gauge boson $W'$ can be
written as:
\begin{equation}
\hat{\sigma}(\hat{s})=\int d\hat{t}\frac{d\hat{\sigma}}{d\hat{t}}
\end{equation}
with
\begin{equation}
\frac{d\hat{\sigma}}{d\hat{t}}=\frac{\pi\alpha^{2}}{4S_{W}^{4}}
[\frac{1}{(\hat{t}-M_{W}^{2})^{2}}+\frac{2g_{L}^{W'qq'}g_{L}^{W'e\nu}}
{(\hat{t}-M_{W}^{2})(\hat{t}-M_{W'}^{2})}+\frac{(g_{L}^{W'qq'}g_{L}^{W'e\nu})^{2}}{(\hat{t}-M_{W'}^{2})^{2}}],
\end{equation}
and $\hat{t}=(P_{1}-P_{4})^{2}$. In above equations, we have assumed
that $W'$ is the pure left-handed charged gauge boson.

The process $eq\rightarrow \nu q'$ can be seen as the subprocess of
the charged current $(CC)$ process $ep\rightarrow \nu q' +X$.
Measurement and $QCD$ analysis of the production cross section for
the $SM$ $CC$ process $ep\rightarrow \nu q' +X$ at the $HERA$
collider have been extensively studied [23]. Including the
contributions of the $SM$ gauge boson $W$ and new gauge boson $W'$,
the production cross section $\sigma_{T}(S)$ of the $CC$ process
$ep\rightarrow \nu q'+X$ at the $ep$ colliders can be written as:
\begin{equation}
\sigma_{T}(S)=\sum_{q}\int^{1}_{x_{\min}}f_{q}(x,\mu)\hat{\sigma}(\hat{s})dx
\end{equation}
with $x_{\min}=m_{q'}^{2}/S$ and $\hat{s}=x S$, in which the
center-of-mass $(c.m.)$ energy $\sqrt{S}$ is taken as 320$GeV$ for
the $HERA$ collider and as 1$TeV$ for the $THERA$ collider. $q$
represents the quarks $u$, $c$, $d$, or $s$. In our numerical
estimation, we will use $CTEQ$6$L$ parton distribution function
$(PDF)$ [24] for the quark distribution function $f_{q}(x,\mu)$ and
assume that the factorization scale $\mu$ is of order
$\sqrt{\hat{s}}$. To take into account detector acceptance, the
angle of the observed jet, $\theta_{q'}$, will be restricted to the
range $10^{\circ}\leq\theta_{q'}\leq170^{\circ}$ [23].

It has been shown [7] that in suitable kinematic region the process
$e\gamma\rightarrow \nu q'\bar{q}$ can be approximated quite well by
the process $eq\rightarrow\nu q'$, where the quark $q$ described by
the quark parton content of the photon approach [25]. The hard
photon beam of $e\gamma$ collision can be obtained from laser
backscattering at the high energy $e^{+}e^{-}$ collider experiments.
The expression for the effective cross section of the subprocess
$eq\rightarrow \nu q'$ at the $ILC$ is given by
\begin{equation}
\sigma_{I}=\sum_{q}\int
dx_{1}dx_{2}f_{\gamma/e}(x_{1})f_{q/\gamma}(x_{2})\hat{\sigma}(\hat{s}),
\end{equation}
where $f_{\gamma/e}(x_{1})$ is the photon distribution [26],
$f_{q/\gamma}$ is the distribution function for the quark content in
the photon. To obtain our numerical results we will use Aurenche,
Fontannaz and Guillet $(AFG)$ distribution [27] for $f_{q/\gamma}$.
Other distributions are available in [28].

In the following sections, we will discuss possibility of detecting
the new charged gauge boson $W'$ in future $THERA$ and $ILC$
experiments via considering its contributions to the subprocess
$eq\rightarrow \nu q'$ in different new physics scenarios.

\noindent{\bf 3. The subprocess \textbf{$eq\rightarrow \nu q'$} in
the little Higgs models}

According to the structure of the extended electroweak gauge group,
the little Higgs models can be generally divided into two classes
[17, 18]: product group models, in which the $SM$ $SU(2)_{L}$ is
embedded in a product gauge group, and simple group models, in which
it is embedded in a large simple group. The $LH$ model [19] and the
$SU(3)$ simple group model [18, 20] are the simplest examples of the
product group models and the simple group models, respectively. To
predigest our calculation, we will discuss the subprocess
$eq\rightarrow \nu q'$ in the context of these two simplest models.

In the $LH$ model, the coupling constants of the $SM$ gauge boson
$W$ and the new gauge boson $W_{H}$ to the first and second
generation fermions, which are related to our calculation, can be
written as [29]:
\begin{equation}
g_{L}^{Wqq'}=\frac{ie}{\sqrt{2}S_{W}}[1-\frac{\nu^{2}}{2f^{2}}c^{2}(c^{2}-s^{2})],\hspace*{0.5cm}
g_{R}^{Wqq'}=0;
\end{equation}

\begin{equation}
\hspace*{-2.7cm}g_{L}^{W_{H}qq'}=\frac{ie}{\sqrt{2}S_{W}}\frac{c}{s},\hspace*{0.5cm}
g_{R}^{W_{H}qq'}=0.
\end{equation}
Here $\nu\approx246GeV$ is the electroweak scale, $c$
$(s=\sqrt{1-c^{2}})$ is the mixing parameter between the $SU(2)_{1}$
and $SU(2)_{2}$ gauge bosons, and $f$ is the scale parameter of the
gauge symmetry breaking.
\begin{figure}[htb]
\begin{center}
\vspace*{0.2cm}
 \epsfig{file=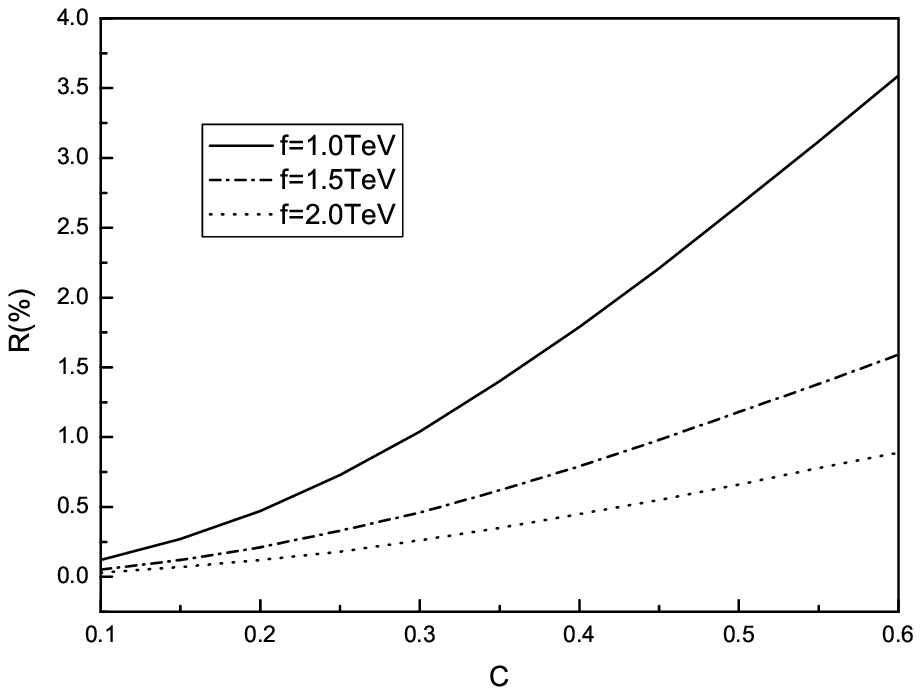,width=225pt,height=210pt}
\put(-110,3){ (a)}\put(115,3){ (b)}
 \hspace{0cm}\vspace{-0.25cm}
\epsfig{file=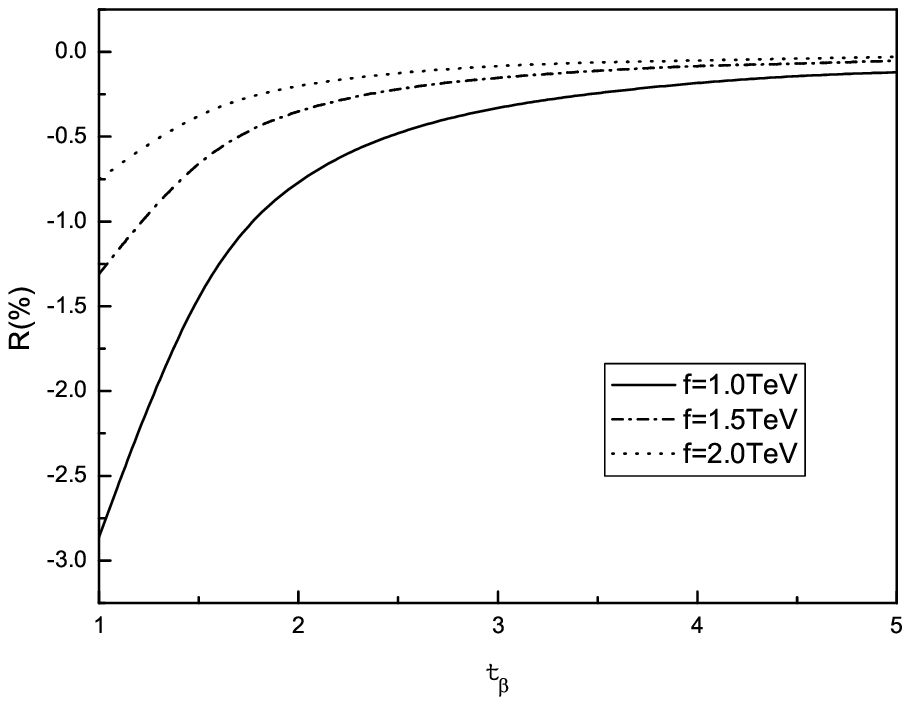,width=225pt,height=210pt} \hspace{-0.5cm}
 \hspace{10cm}\vspace{-1cm}
 \caption{At the $THERA$, the relative correction parameter R as
 function of the mixing  \hspace*{1.8cm}parameter $c$ for the $LH$ model (a) and of the
parameter $t_{\beta}$ for the $SU(3)$ simple \hspace*{1.8cm}group
model (b) for three  values of the scale parameter
$f$.\vspace{0.05cm}}
 \label{ee}
\end{center}
\end{figure}

Similar with the $LH$ model, the $SU(3)$ simple group model [18, 20]
also predicts the existence of the new charged gauge boson, which is
represented by $X$. In the $SU(3)$ simple group model, the coupling
constants of the $SM$ gauge boson $W$ and the new gauge boson $X$ to
the first and second generation fermions can be written as:
\begin{equation}
g_{L}^{Wqq'}=\frac{ie}{\sqrt{2}S_{W}}(1-\frac{1}{2}\delta_{\nu}^{2}),\hspace*{0.5cm}
g_{R}^{Wqq'}=0;
\end{equation}
\begin{equation}
\hspace*{-1.5cm}g_{L}^{Xqq'}=\frac{ie}{\sqrt{2}S_{W}}\delta_{\nu},\hspace*{0.5cm}
g_{R}^{Xqq'}=0
\end{equation}
with $\delta_{\nu}=-\nu/2ft_{\beta}$.
 Here
$f=\sqrt{f_{1}^{2}+f_{2}^{2}}$ and
$t_{\beta}=\tan\beta=f_{2}/f_{1}$, in which $f_{1}$ and $f_{2}$ are
the vacuum condensate values of the two sigma-model fields
$\Phi_{1}$ and $\Phi_{2}$, respectively.

After taking into account electroweak symmetry breaking $(EWSB)$, at
the leading order, the masses of the new charged gauge bosons
$W_{H}$ and $X$ can be written as:
\begin{equation}
M_{W_{H}}=\frac{gf}{2sc},\hspace*{0.5cm}M_{X}=\frac{gf}{\sqrt{2}}.
\end{equation}

Except for the $SM$ input parameters $\alpha=1/128.8$,
$S_{W}^{2}=0.2315$, and $M_{W}=80.14GeV$ [8], the contributions of
the $LH$ model and the $SU(3)$ simple group model to the production
cross section of the subprocess $eq\rightarrow \nu q'$ dependent on
the free parameters $(f,c)$ and $(f,t_{\beta})$, respectively.
Considering the constraints of the electroweak precision data on
these free parameters, we will assume $1TeV\leq f\leq 3TeV$ and
$0<c\leq0.6$ for the $LH$ model [30], and $1TeV\leq f\leq3TeV$ and
$t_{\beta}>1$ for the $SU(3)$ simple group model [17, 18, 20] in our
numerical estimation.

\begin{figure}[htb]
\begin{center}
\vspace{0.2cm}
 \epsfig{file=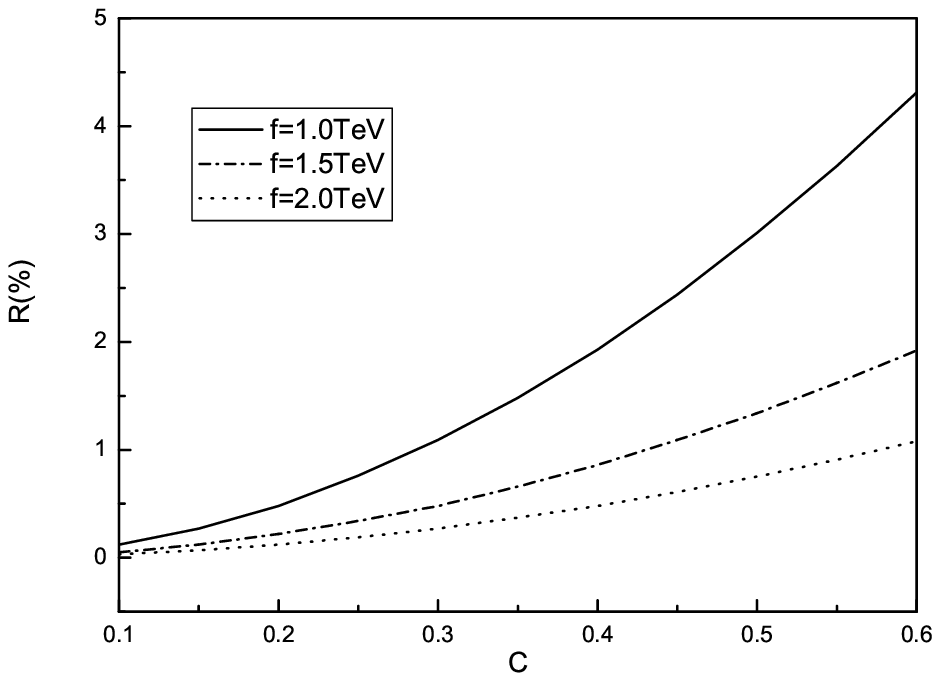,width=225pt,height=210pt}
\put(-110,3){ (a)}\put(115,3){ (b)}
 \hspace{0cm}\vspace{-0.25cm}
\epsfig{file=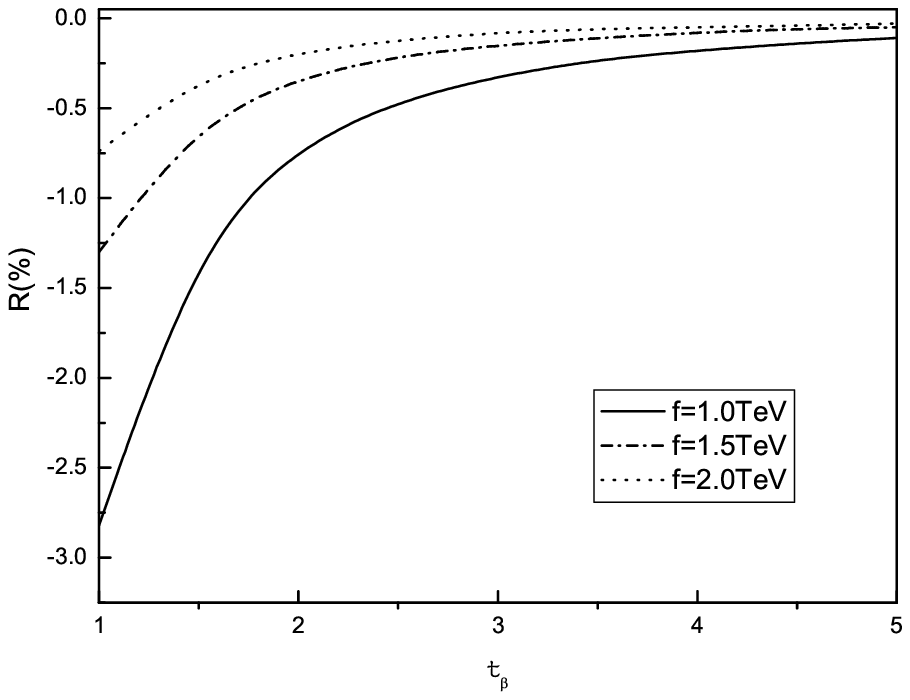,width=225pt,height=210pt} \hspace{-0.5cm}
 \hspace{10cm}\vspace{-1cm}
 \caption{Same as Fig.1 but for $ILC$.}
 \label{ee}
\end{center}
\end{figure}

\vspace{-0.5cm}
 To illustrate the contributions of the new physics model to the
 subprocess $eq\rightarrow \nu q'$, we define the relative
 correction parameter $R$=$\frac{\sigma(i)-\sigma(SM)}{\sigma(SM)}$,
 in which $\sigma(i)$ and $\sigma(SM)$ represent the effective cross
 sections predicted by the new physics model and the $SM$,
 respectively. The relative correction parameters for the $LH$ model and the $SU(3)$
 simple group model at the $THERA$ and $ILC$ experiments are
plotted in Fig.1 and Fig.2, respectively. In these figures, we have
assumed the $CKM$ matrix elements $V_{ud}\approx V_{cs}\approx1$ and
taken the $c.m.$ energy $\sqrt{S}=1000GeV$ and 500$GeV$ for the
$THERA$ and $ILC$ experiments, respectively. One can see from these
figures that the $LH$ model can give positive contributions to the
effective cross sections at the $THERA$ and $ILC$ experiments, while
the $SU(3)$ simple group model can give negative contributions. The
absolute value of the relative correction parameter $R$ for the
$SU(3)$ simple group model is slight smaller than that for the $LH$
model. For the $SU(3)$ simple group model, the values of $R$ at the
$THERA$ are approximately equal to those at the $ILC$. However, in
most of the parameter spaces for the $LH$ model and the $SU(3)$
simple group model, all of the absolute values of the relative
correction parameter $R$ are smaller than $4.3\%$ .

\begin{figure}[htb]
\begin{center}
\vspace{0.2cm}
 \epsfig{file=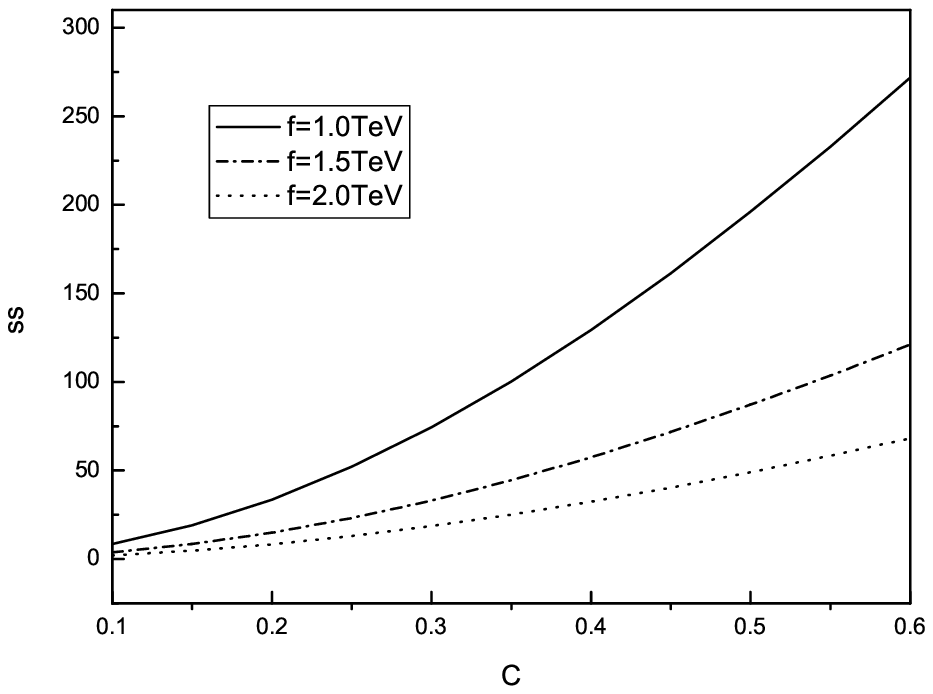,width=225pt,height=210pt}
\put(-110,3){ (a)}\put(115,3){ (b)}
 \hspace{0cm}\vspace{-0.25cm}
\epsfig{file=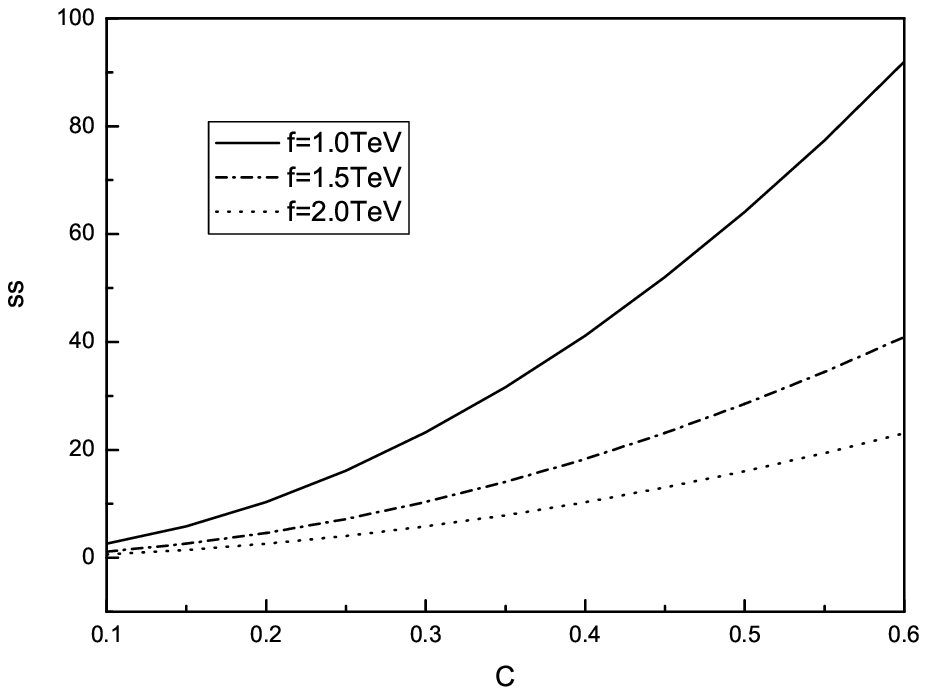,width=225pt,height=210pt} \hspace{-0.5cm}
 \hspace{10cm}\vspace{-1cm}
 \caption{For the $LH$ model, $SS$ as a function of the mixing parameter $c$
 for three values \hspace*{1.8cm}of $f$ at the $THERA$ (a) and the $ILC$ (b).}
 \label{ee}
\end{center}
\end{figure}

\begin{figure}[htb]
\begin{center}
\vspace{0.2cm}
 \epsfig{file=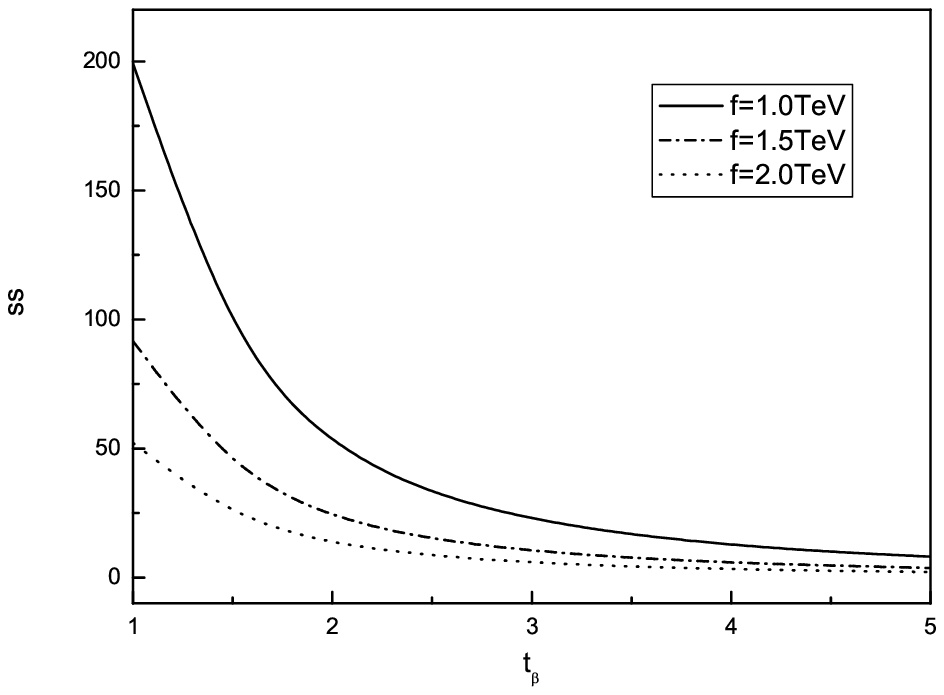,width=225pt,height=210pt}
\put(-110,3){ (a)}\put(115,3){ (b)}
 \hspace{0cm}\vspace{-0.25cm}
\epsfig{file=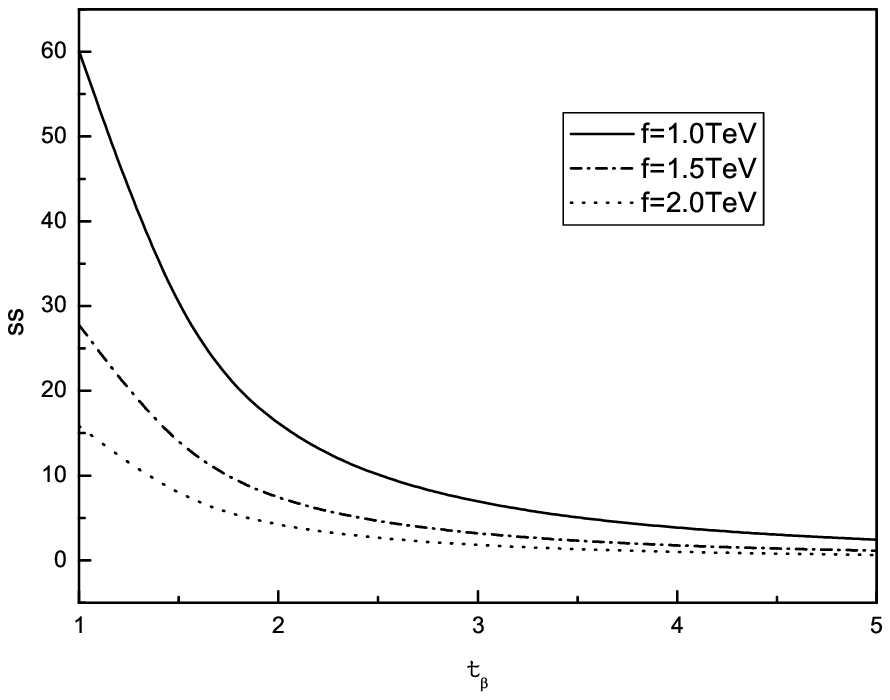,width=225pt,height=210pt} \hspace{-0.5cm}
 \hspace{10cm}\vspace{-1cm}
 \caption{For the $SU(3)$ simple group model, $SS$ as a function of the parameter $t_{\beta}$
 for \hspace*{1.87cm}three values of $f$ at the $THERA$ (a) and the $ILC$ (b).}
 \label{ee}
\end{center}
\end{figure}

\vspace{-0.5cm} In order to see if the correction effects of the
$LH$ model and the $SU(3)$ simple group model on the processes
$ep\rightarrow \nu q'+X$ and $e^{+}e^{-}\rightarrow \nu q'+X$ can be
observed in the future $THERA$ and $ILC$ experiments, we define the
statistical significance $(SS)$ of the signal as:
\begin{equation}
SS^{i}=\frac{\mid
\sigma(i)-\sigma(SM)\mid}{\sqrt{\sigma(SM)}}\sqrt{\pounds}.
\end{equation}
Here $i$ represents the $LH$ model or the $SU(3)$ simple group
model. In our numerical calculation, we will assume the values of
yearly integrated luminosity $\pounds$ as $4fb^{-1}$ and
$100fb^{-1}$ for the $THERA$ experiment with $ \sqrt{S}=1000GeV$ and
$ILC$ experiment with $ \sqrt{S}=500GeV$, respectively. Our
numerical results are summarized in Fig.3 and Fig.4. One can see
from these figures that, for these two little Higgs models, the
value of $SS$ at the $THERA$ is larger than that at the $ILC$. For
the assumed integrated luminosity, the effects of the little Higgs
models on the subprocess $eq\rightarrow\nu q'$ can generally be
easier detected at the $THERA$ than at the $ILC$. For same high
energy collider experiment ($ILC$ or $THERA$), the $SS$ value
contributed by the $LH$ model is larger than that by the $SU(3)$
simple group model.  For the $ILC$ experiment with $\sqrt{S}=500GeV$
and $\pounds=100fb^{-1}$, if we take $f=2TeV$, $0.2\leq c\leq 0.6$
and $1\leq t_{\beta}\leq 2.5$, the values of $SS$ are in the ranges
of $2.6\sim 23.1$ and $15.8\sim 2.5$ for the $LH$ model and the
$SU(3)$ simple group model, respectively. Thus, with reasonable
values of the free parameters, the possible signatures of the new
charged gauge boson $W'$ predicted by the $LH$ model or by the
$SU(3)$ simple group model might be detected via the subprocess
$eq\rightarrow \nu q'$ in the future $ILC$ and $THERA$ experiments.

\noindent{\bf 4. The subprocess \textbf{$eq\rightarrow \nu q'$} in
the three-site Higgsless model}

So far, various kinds of models for $EWSB$ have been proposed, among
which Higgsless model [2] is one of the attractive new physics
models. In this kind of models, $EWSB$ can be achieved via employing
gauge symmetry breaking by boundary condition in higher dimensional
theory space [31], and the unitary of longitudinally polarized $W$
boson and $Z$ boson scattering is preserved by exchange of new
vector gauge bosons [32]. Reconstructed Higgless models [15, 16]
have been used as tools to compute the general properties of
Higgsless models and to illustrate the phenomenological properties
of this kind of new physics models beyond the $SM$.

The simplest deconstructed Higgsless model incorporates only three
sites on the deconstructed lattice, which is called the three-site
Higgsless model [16]. In this model, the ordinary fermions are
ideally delocalized, which preserves the characteristic of vanishing
precision electroweak corrections up to subleading order [33].
Furthermore, the three-site Higgsless model is capable of
approximating much of the interesting phenomenology associated with
extra dimensional models and more complicated deconstructed
Higgsless models [34].

The three-site Higgsless model [16] has a standard color group and
an extended $SU(2)_{1}\times SU(2)_{2}\times U(1)$ electroweak gauge
group, which is similar to that of the $BESS$ model [35]. Once
$EWSB$ occurs in this model, the gauge sector consists of a massless
photon, two relatively light massive gauge bosons which are
identified with the $SM$ $W$ and $Z$ gauge bosons, as well as two
heavy gauge bosons which are denoted as $Z'$ and $W'$. In the
three-site Higgsless model, the coupling constants of the charged
gauge bosons $W$ and $W'$ to ordinary fermions can be written as:
\begin{equation}
g_{L}^{Wff'}=\frac{iS_{W}}{e}[g(1-x_{1})a_{22}+\widetilde{g}x_{1}a_{12}],\hspace*{0.5cm}
g_{R}^{Wff'}=0;
\end{equation}
\begin{equation}
\hspace*{0.1cm}g_{L}^{W'ff'}=\frac{iS_{W}}{e}[g(1-x_{1})a_{21}+\widetilde{g}x_{1}a_{11}],\hspace*{0.5cm}
g_{R}^{W'ff'}=0.
\end{equation}
Here the parameter $x_{1}$ is a measure of the amount of fermion
delocalization $(0<x_{1}\ll1)$ [16, 33]. In principle, the value of
$x_{1}$ for a given fermion species depends indirectly on the mass
of the fermion. However, since we are only interested in light
fermions, we can assume that the parameter $x_{1}$ has the same
value for the first- and second- generation fermions. The expression
forms of the parameters $g$, $\tilde{g}$, $a_{22}$, $a_{12}$,
$a_{21}$, and $a_{11}$ have been given by [36] in terms of the $W$
and $W'$ masses $M_{W}$ and $M_{W'}$. In our numerical estimation,
we will assume $M_{Z'}^{2}=M_{W'}^{2}+(M_{Z}^{2}-M_{W}^{2})$, and
take $x_{1}$ and $M_{W'}$ as free parameters.

\begin{figure}[htb]
\begin{center}
\vspace{0.2cm}
 \epsfig{file=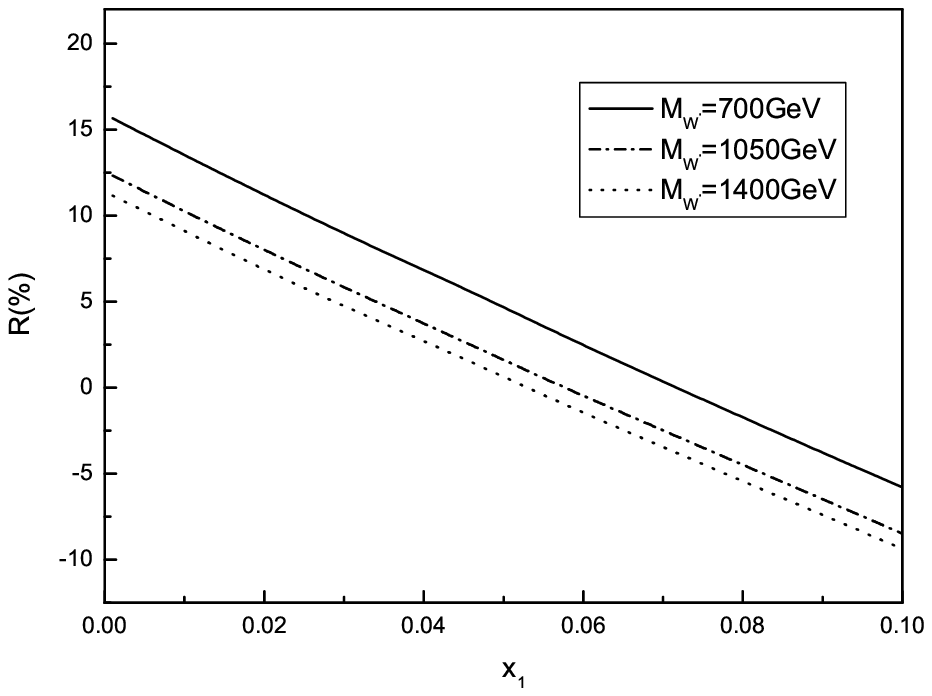,width=225pt,height=210pt}
\put(-110,3){ (a)}\put(115,3){ (b)}
 \hspace{0cm}\vspace{-0.25cm}
\epsfig{file=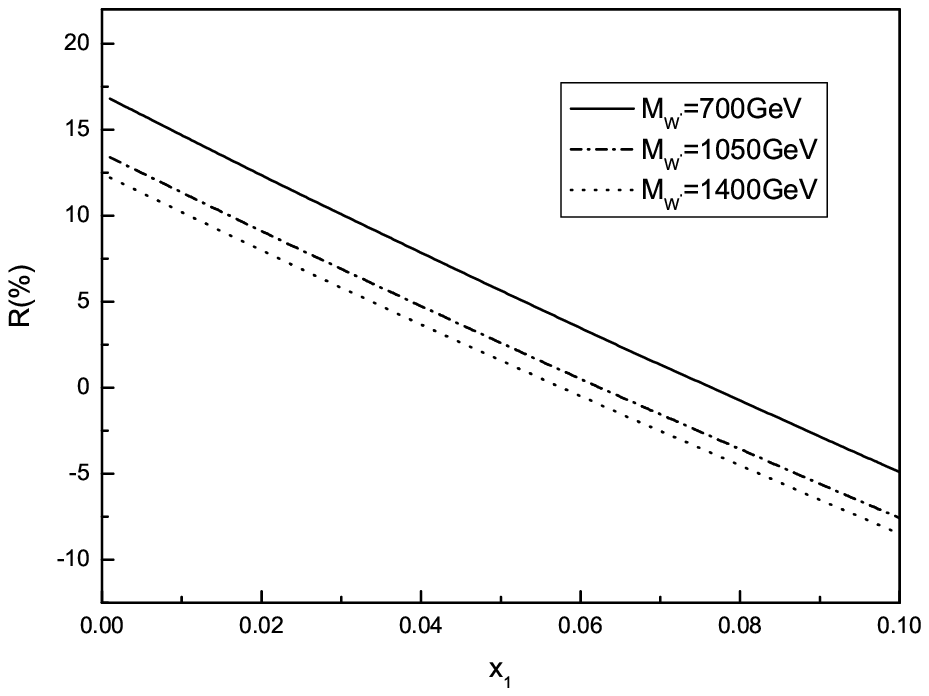,width=225pt,height=210pt} \hspace{-0.5cm}
 \hspace{10cm}\vspace{-1cm}
 \caption{The relative correction parameter
 $R$ varies
 as the parameter $x_{1}$ for three values  \hspace*{1.87cm}of the $W'$ mass $M_{W'}$
 at the $THERA$ (a) and the $ILC$ (b).}
 \label{ee}
\end{center}
\end{figure}

Our numerical results obtained in the content of the three-site
Higgsless model are given in Fig.5 and Fig.6, in which we have
assumed $M_{W'}$=$700GeV$, 1050$GeV$, and $1400GeV$. One can see
from these figures that the contributions of the three-site
Higgsless model  to the subprocess $eq\rightarrow \nu q'$ depend
rather significantly on the free parameter $x_{1}$. The value of the
relative correction parameter $R$ is positive or negative, which
depends on the value of the free parameter $x_{1}$. The value of $R$
for the $ILC$ experiment with $\sqrt{S}=500GeV$ is approximately
equal to that for the $THERA$ experiment with $\sqrt{S}=1TeV$.
However, the statistical significance $SS$ of the signal for the
$THERA$ experiment is larger than that for the $ILC$ experiment. In
wide range of the parameter space, the value of $SS$ is
significantly large. Thus, we expect that the correction effects of
the three-site Higgsless model to the subprocess $eq\rightarrow \nu
q'$ can be observed in the future $THERA$ and $ILC$ experiments.

\begin{figure}[htb]
\begin{center}
\vspace{0.2cm}
 \epsfig{file=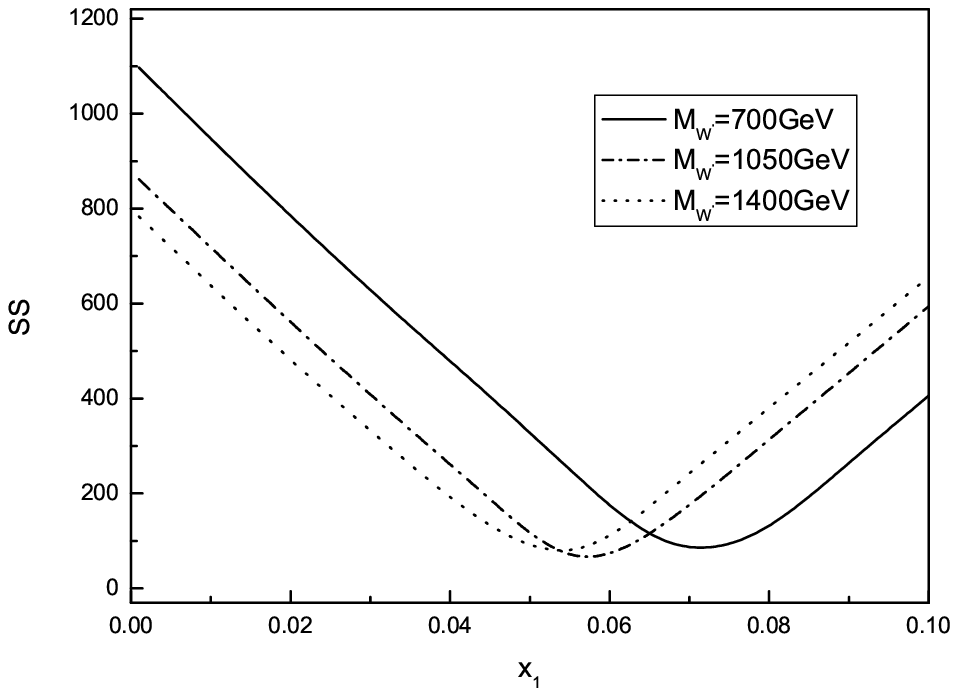,width=225pt,height=210pt}
\put(-110,3){ (a)}\put(115,3){ (b)}
 \hspace{0cm}\vspace{-0.25cm}
\epsfig{file=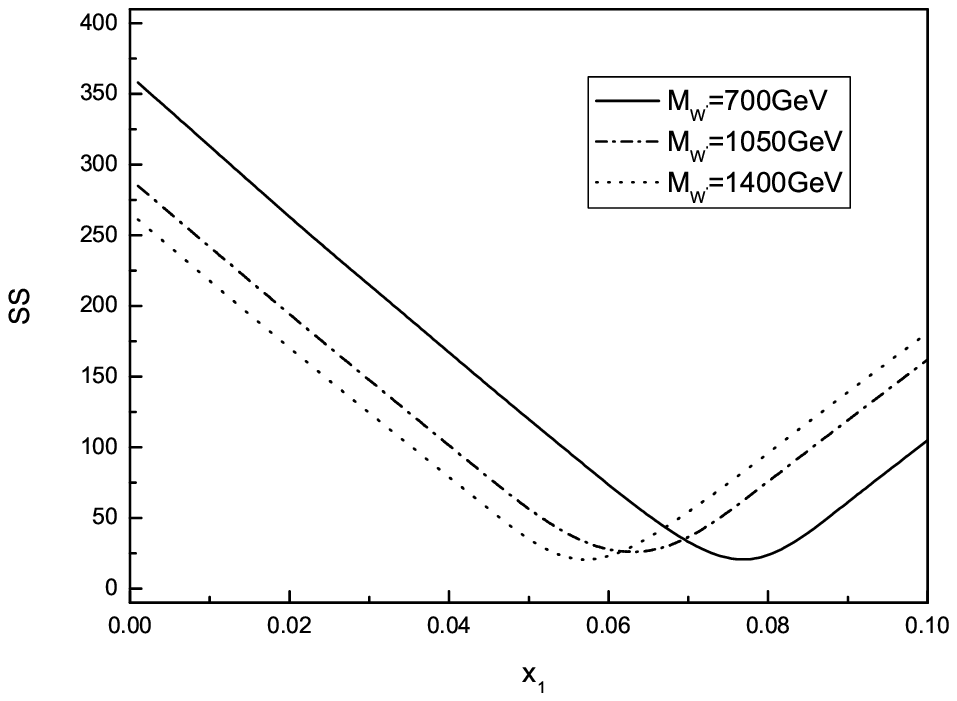,width=225pt,height=210pt} \hspace{-0.5cm}
 \hspace{10cm}\vspace{-1cm}
 \caption{$SS$ as a function of the parameter $x_{1}$ for three
 values of the $W'$ mass $M_{W'}$ at  \hspace*{1.87cm}the $THERA$ (a) and the $ILC$ (b).}
 \label{ee}
\end{center}
\end{figure}

 \noindent{\bf 5. Conclusions and discussions }

Most of all the new physics models beyond the $SM$ predict the
existence of the new charged gauge boson $W'$, which might generate
observed signatures in future high energy collider experiments. The
$W'$ arised from different new physics models can induce different
physical signatures. Thus, it is very interesting to study the
correction effects of the new gauge boson $W'$ on some observables.
It will be helpful to test the $SM$ and further to distinguish
different new physics models.

The process $eq\rightarrow \nu q'$ mediated by the charged gauge
boson $W'$ can be seen as the subprocess of the processes
$ep\rightarrow\nu q'+X$ and $e^{+}e^{-}\rightarrow \nu q'+X$. One
can use the subprocess $eq\rightarrow \nu q'$ to detect the possible
signals of the new charged gauge boson $W'$ in future $THERA$ and
$ILC$ experiments. Ref.[39] has studied the contributions of the
four fermion contact terms to this subprocess. In this paper, we
study the contributions of the $W'$ predicted by the little Higgs
models and the three-site Higgslesss model to this subprocess and
discuss the possibility of detecting $W'$ in future $THERA$ and
$ILC$ experiments. Our numerical results are summarized in Table 1.
\begin{table}[hbt]
\begin{center}
\caption{The contributions of the $LH$ model, $SU(3)$ simple group
model, and the three-\hspace*{1.7cm}site Higgslesss model to the
subprocess $eq\rightarrow \nu q'$ at the $THERA$ and $ILC$
\hspace*{1.7cm}experiments.}\vspace{-0.5cm}
\tabcolsep=1pt%
\begin{tabular}{|c|c|c|c|}
\hline
{\hbox{Models}} &\hspace{-0.5cm}{\hbox{$LH$}}&\hspace{-0.9cm}{\hbox{$SU(3)$}}&\hspace{-0.8cm}{\hbox{$HL$}}\\
\hline  $$& \hspace{-0.2cm}$f=2TeV$  & \hspace{-0.2cm}$f=2TeV$&
\hspace{-0.2cm}$M_{W'}=1050GeV$
\\
$$ &\hspace{-0.2cm}$ 0.2\leq c\leq 0.6$ &\hspace{-0.2cm}$ 1\leq t_{\beta}\leq2.5$& \hspace{-0.2cm}$ 0.002\leq
x_{1}\leq0.08$
\\
\cline{2-2}\cline{3-3}\cline{4-4} \hline$R$$(\%)$&
   \begin{tabular}{c|c} \hspace{-0.3cm}\makebox[1.7cm][s]{$THERA$}&\hspace{0.6cm}$ILC$
\end{tabular}
& \begin{tabular}{c|c} \hspace{-0.9cm}\makebox[2.3cm][s]{$THERA$}
&\hspace{0.6cm}$ILC$
\end{tabular}
& \begin{tabular}{c|c}
\hspace{-0.2cm}\makebox[2cm][s]{$THERA$}&\hspace{0.6cm}$ILC$
\end{tabular}\\
\cline{2-2}\cline{3-3}\cline{4-4}
$$
&
\begin{tabular}{c|c}\hspace{-0.0cm}$0.12\sim0.89$
&$0.12\sim1.08$
\end{tabular}
&\begin{tabular}{c|c} \hspace{-0.0cm}$-0.75\sim-0.12$ &
$-0.74\sim-0.12$
\end{tabular}
& \begin{tabular}{c|c} \hspace{-0.0cm}$12.1\sim-4.5$ &
$13.2\sim-3.6$
\end{tabular}\\
 \hline$SS$&
\begin{tabular}{c|c} \hspace{-0.0cm}$8.4\sim68.1$& $2.6\sim23.1$
\end{tabular}
& \begin{tabular}{c|c} \hspace{-0.0cm}$52.2\sim8.4$ & $15.8\sim2.5$
\end{tabular}
& \begin{tabular}{c|c} \hspace{0.2cm}$845.5\sim314.2$ &
$280.8\sim76.1$
\end{tabular}\\
\hline
\end{tabular}
\end{center}
\end{table}

The contributions of the three-site Higgsless model to the
subprocess $eq\rightarrow \nu q'$ are generally larger than those
for the $LH$ model or the $SU(3)$ simple group model. The effects of
the three-site Higgsless model on this subprocess can generally be
easier detected than those for the little Higgs models. However, it
can enhance or reduce the effective cross sections of the subprocess
$eq\rightarrow \nu q'$ at the $THERA$ and $ILC$ experiments, which
depends on the value of the free parameter $x_{1}$. Thus, we can use
the subprocess $eq\rightarrow \nu q'$ to detect the possible
signatures of the new charged gauge boson $W'$ and further
distinguish the three-site Higgsless model and the little Higgs
models in future $THERA$ or $ILC$ experiments.

In this paper, we have assumed that the hard photon beam is obtained
from laser backscattering. Certainly, we can also take that the hard
photon beam arises from Weizs$\ddot{a}$cker Williams bremsstrahlung
[37]. Furthermore, in our numerical estimation, we have taken $AFG$
$PDFs$ for the quark distribution functions in the photon. Other
$PDFs$ can also be used to give our numerical results. These will
change the above numerical results. However, they can not change our
physical conclusions.

In order to satisfy the electroweak precision constraints by
avoiding tree-level contributions of the new particles and restoring
the custodial $SU(2)$ symmetry, a discrete symmetry (called
T-parity) is introduced to the $LH$ model, which forms the so called
$LHT$ model [38].  Under T-parity, particle fields predicted by this
model are divided into T-even and T-odd sectors. The T-even sector
consists of the $SM$ particles and a heavy top $T_{+}$, while the
T-odd sector contains heavy gauge bosons $(B_{H}, Z_{H},
W_{H}^{\pm})$, a scalar triplet $(\Phi)$, and the so-called mirror
fermions $(L_{H}, Q_{H})$. The mirror quark can be produced via the
process $eq\rightarrow \nu_{H}Q_{H}$ mediated by the T-odd charged
gauge boson $W_{H}$, which can give similar signal with that from
the process $eq\rightarrow \nu q'$. We will study the process
$eq\rightarrow \nu_{H}Q_{H}$ in near future works. \vspace{1.0cm}

\noindent{\bf Acknowledgments}

This work was supported in part by the National Natural Science
Foundation of China under Grants No.10675057 and Foundation of
Liaoning  Educational Committee(2007T086).

\end{document}